\documentclass[aps,prd,twocolumn, showpacs, showkeys]{revtex4-1}

\usepackage{graphicx}

\begin{document}

\title{Modified arctan-gravity model mimicking a cosmological constant}

\author{S. I. Kruglov}

\affiliation{Department of Chemical and Physical Sciences, University of Toronto,\\
3359 Mississauga Road North, Mississauga, Ontario, Canada L5L 1C6}

\begin{abstract}
A novel theory of $F(R)$ gravity with the Lagrangian density ${\cal L}=[R-(b/\beta)\arctan\left(\beta R\right)]/(2\kappa^2)$ is analyzed. Constant curvature solutions of the model are found, and the potential of the scalar field and the mass of a scalar degree of freedom in Einstein's frame are derived. The cosmological parameters of the model are calculated, which are in agreement with the PLANCK data. Critical points for the de Sitter phase and the matter dominated epoch of autonomous equations are obtained and studied.
\end{abstract}


\maketitle

\section{Introduction}

Inflationary cosmology, introduced by Guth \cite{Guth}, remains the main point of view in modern cosmology.
It can solve the problem of the initial conditions to explain the formation of galaxies and irregularities in the microwave background. Therefore, it is important to suggest the model that describes correctly the inflationary epoch in accordance with the experimental data. The evolution of the Universe can be characterized by the cosmological parameters that were measured by WMAP and recent PLANCK experiments, allowing us to select the viable model from the many models suggested. The $\Lambda$CDM ($\Lambda$-Cold Dark Matter) model introducing the cosmological constant $\Lambda$ \cite{Frieman} remains  a good candidate for a description of dark energy (DE) and all observational data of the accelerated expansion. But this model suffers difficulty with the explanation of the smallness of $\Lambda$ compared with the vacuum energy of elementary particles. To describe the accelerated Universe at the present time the single-field model with dynamical dark energy may be used \cite{Linde}. Here we pay attention to $F(R)$-gravity theories and the modification of the general relativity (GR), replacing the Ricci scalar in Einstein-Hilbert action by the function $F(R)$ \cite{Caldwell}, \cite{Faraoni}, \cite{Capozziello}, \cite{Odintsov}. $F(R)$-gravity models contain a single extra degree of freedom (a scalar) and can be reformulated in a scalar-tensor form. Such purely gravitational models describe the inflation, modifies gravitational physics, and are an alternative to the $\Lambda$CDM model. The first successful models of $F(R)$ gravity were given in \cite{Starobinsky}, \cite{Hu}, \cite{Appl}, \cite{Star}. Some models of $F(R)$-gravity theories were considered in
\cite{Deser}, \cite{Nojiri}, \cite{Carroll}, \cite{Amendola}, \cite{Cognola}, \cite{Linder}, \cite{Ferreira}, \cite{Pani}, \cite{Kruglov}, \cite{Kruglov1}, and in many other publications. The requirement is that the scalar field is not a tachyon and a ghost leads to the condition $F''(R)>0$ (primes mean derivatives with respect to an argument). At the same time the inequality $F'(R)>0$ guarantees that a graviton is not a ghost \cite{Star}. The $F(R)$ gravity is phenomenological effective model that describes geometrical DE. If the $F(R)$-gravity model really describes the evolution of our Universe, the present and primordial DE, it should follow from the fundamental theory like string or M theory.\\
In this paper we consider the choice of the function corresponding to $F(0) = 0$, $F(R)\rightarrow R-2\Lambda_{eff}$ at $R\gg\Lambda_{eff}$, where $\Lambda_{eff}$ is the effective cosmological constant. The $\Lambda_{eff}$ is not zero in curved space-time but is zero in the flat space-time. Such a model mimics the cosmological constant at large $R$ due to geometry, and $\Lambda_{eff}$ is not connected with quantum vacuum energy. Such a class of models was already considered in \cite{Star}, \cite{Hu}, \cite{Appl}. Here we suggest another $F(R)$-gravity model that passes cosmological tests.

The paper is organized as follows. In Sec. 2 we formulate the model showing the classical and quantum stabilities. The constant curvature solutions of equations of motion are obtained and the de Sitter phases are considered. After performing the conformal transformation of the metric we obtain the scalar-tensor form of the model in Einstein's frame in Sec. 3. The potential of the scalar field and the mass of a scalaron are derived and the plots of the functions $\phi(\beta R)$, $V(\beta R)$, $V(\phi)$, and $m^2_{\phi}(\beta R)$ for different values of the parameter $b$ are given. In Sec. 4 the cosmological slow-roll parameters of the model are evaluated, and the plots of $\epsilon(\beta R)$, $\eta(\beta R)$, and $n_s(\beta R)$ are presented. In Sec. 5 critical points of autonomous equations are investigated showing that the matter dominated epoch is viable. The function $m(r)$ characterizing the deviation from the
$\Lambda$CDM model is calculated and the plot is given. The effective gravitational constant related to the evolution of matter density perturbations is studied. Section 6 is devoted to the conclusion.

The Minkowski metric $\eta_{\mu\nu}$=diag(-1, 1, 1, 1) is used, and  $c=\hbar=1$ is assumed throughout the paper.

\section{The model}

Let us consider the gravitation theory where we replace the Ricci scalar $R$ in the Einstein-Hilbert action by the function
\begin{equation}
F(R)=R+f(R),~~~~f(R)=-\frac{b}{\beta}\arctan\left(\beta R\right),
\label{1}
\end{equation}
and we imply that $\beta >0$, $R>0$. The constant $\beta$ has the dimension of (length)$^2$ and $b$ is the dimensionless constant. The function $F(R)$ satisfies the conditions $F(0)=0$, $\lim_{R\rightarrow\infty}f(R)=\mbox{const}$ which are necessary to mimic the phenomenology of the $\Lambda$CDM model in the high-redshift regime. Thus, the action in the Jordan frame becomes
\begin{equation}
S=\int d^4x\sqrt{-g}{\cal L}=\int d^4x\sqrt{-g}\left[\frac{1}{2\kappa^2}F(R)+{\cal L}_{\mbox{m}}\right],
\label{2}
\end{equation}
where $\kappa=M_{Pl}^{-1}$, $M_{Pl}$ is the reduced Planck mass, and ${\cal L}_{\mbox{m}}$ is the matter Lagrangian density. From Eq.(1) one obtains
\begin{equation}
F'(R)=1-\frac{b}{1+(\beta R)^2},~~~~F''(R)=\frac{2b\beta^2 R}{\left(1+(\beta R)^2\right)^2},
\label{3}
\end{equation}
where primes denote the derivatives on argument. The function $F(R)$ obeys the condition $F''(R)>0$ at $b>0$, which ensures the classical stability of the solution at high curvature. It follows from Eq.(3) that the condition of quantum stability $F'(R)>0$ leads to
\begin{equation}
1+(\beta R)^2>b.
\label{4}
\end{equation}
Thus, $0<b<1$.

\subsection{Constant curvature solutions}

Let us consider constant curvature solutions to equations of motion that follow from the action (2) without matter. Such equation of motion is given by \cite{Barrow}
\begin{equation}
2F(R)-RF'(R)=0.
\label{5}
\end{equation}
It should be noted that Eq.(5) corresponds to the extremum of the effective potential of the scalar degree of freedom (in Einstein's frame). With the help of Eqs. (1) and (5) we obtain
\begin{equation}
b\left(2\arctan (\beta R)-\frac{\beta R}{1+(\beta R)^2}\right)=\beta R.
\label{6}
\end{equation}
Equation (6) possesses the solution $R_0=0$ corresponding to the Minkowski space-time.
Nontrivial solutions of the transcendental equation (6) for the model with $b=0.99$ are given by
\[
\beta R_1\approx 0.1791~~~~(\kappa \phi_1\approx 3.919),
\]
\[
\beta R_2\approx 1.4582~~~~(\kappa \phi_2\approx 0.466),
\]
and for $b=0.999$ are
\[
\beta R_1\approx 0.05495~~~~(\kappa \phi_1\approx 6.760),
\]
\[
\beta R_2\approx 1.5094~~~~(\kappa \phi_2\approx 0.445).
\]
It should be noted that for $b\leq 0.9$ Eq.(6) has only trivial solution $R_0=0$ which goes with flat space-time.
Nontrivial constant curvature solutions $R_1$ and $R_2$ lead to the Schwarzschild-de Sitter space and correspond to inflation. If the condition $F'(R)/F''(R)>R$ is satisfied, then it can describe primordial and present dark energy which are future stable \cite{Schmidt}. From Eq.(3), one obtains
\begin{equation}
\frac{F'(R)}{F''(R)}=\frac{\left[1+(\beta R)^2\right]\left[1+(\beta R)^2-b\right]}{2b\beta^2 R}.
\label{7}
\end{equation}
It can be verified that the constant curvature solutions $R_{1}$ are unstable because
$F'(R_{1})/F''(R_{1})<R_{1}$ and correspond to the maximum of the potential for scalaron (a scalar degree of freedom) and solutions $R_{2}$ are stable as
$F'(R_{2})/F''(R_{2})>R_{2}$ and correspond to the minimum of the potential for scalaron.\\
The spherically symmetric metric of the Schwarzschild form is given by
\begin{equation}
ds^2=-B(r)dt^2+\frac{dr^2}{B(r)}+r^2\left(d\theta^2+\sin^2\theta d\phi^2\right).
\label{8}
\end{equation}
$F(R)$-gravity theories with the constant Ricci scalar $R>0$ have
Schwarzschild$-$ de Sitter solutions with the function $B(r)$,
\begin{equation}
B(r)=1-\frac{2MG}{r}-\frac{R}{12}r^2,
\label{9}
\end{equation}
possessing the classical stability of Schwarzschild black holes; $G$ is the Newton constant and $M$ is the mass of the black hole. Einstein's equation solutions with the cosmological constant $\Lambda$ have the function of the form (9) with $R=4\Lambda$. Thus, the model under consideration possesses the dynamical cosmological constants. Therefore, for the space without any matter, our model mimics DE (the cosmological constant).\\
In the homogeneous, isotropic, and spatially flat FRW (Friedmann-Robertson-Walker) cosmology the Ricci scalar $R$ is given by $R = 12H^2 + 6\dot{H}$, where $H=\dot{a}(t)/a(t)$ is the Hubble parameter. For the case when the Hubble parameter is constant,  $H_0=\sqrt{R/12}$ ($R$ is the constant curvature solution), we have a de Sitter phase. Then a scale factor becomes
$a(t)=a_0\exp (H_0 t)$ ($a_0$ is a scale factor at a cosmic time $t=0$), and this solution describes the  eternal inflation phase.

\section{The scalar-tensor form}

\begin{figure}[h]
\includegraphics[height=2.5in,width=2.9in]{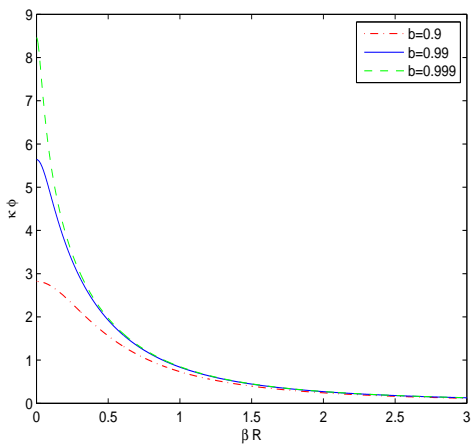}
\caption{\label{fig.1}$\kappa\phi$ versus $\beta R$.}
\end{figure}

\begin{figure}[h]
\includegraphics[height=2.5in,width=2.9in]{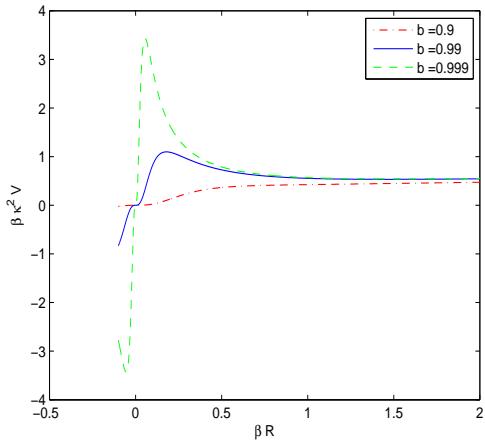}
\caption{\label{fig.2}$\beta \kappa^2 V$ versus $\beta R$.}
\end{figure}

In the Einstein frame, corresponding to the scalar-tensor theory of gravity, we have conformally transformed metric \cite{Sokolowski}
\begin{equation}
\widetilde{g}_{\mu\nu} =F'(R)g_{\mu\nu}=\frac{1-b+(\beta R)^2}{1+(\beta R)^2}g_{\mu\nu}.
\label{10}
\end{equation}
In this frame action (2), at ${\cal L}_m=0$, is given by
\begin{equation}
S=\int d^4x\sqrt{-g}\left[\frac{1}{2\kappa^2}\widetilde{R}-\frac{1}{2}\widetilde{g}^{\mu\nu}
\nabla_\mu\phi\nabla_\nu\phi-V(\phi)\right],
\label{11}
\end{equation}
where $\nabla_\mu$ is the covariant derivative, and $\widetilde{R}$ is calculated with the help of new metric (10). The scalar field $\phi$ and the potential $V(\phi)$ read
\begin{equation}
\phi=-\frac{\sqrt{3}\ln F'(R)}{\sqrt{2}\kappa}=\frac{\sqrt{3}}{\sqrt{2}\kappa}\ln \left(\frac{1+(\beta R)^2}{1-b+(\beta R)^2}\right),
\label{12}
\end{equation}
\begin{widetext}
\begin{equation}
V(\phi) =\frac{RF'(R)-F(R)}{2\kappa^2F'^2(R)}|_{R=R(\phi)}
=b\frac{\left(1+(\beta R)^2\right)\left[\left(1+(\beta R)^2\right)\arctan(\beta R)-\beta R\right]}{2\kappa^2\beta\left[1-b+(\beta R)^2\right]^2}|_{R=R(\phi)},
\label{13}
\end{equation}
\end{widetext}
where the curvature $R(\phi)$ is expressed by Eq. (12) via $\phi$.
The plots of the function $\kappa\phi (R)$ for different parameters $b$ are presented in Fig. \ref{fig.1}.
One can verify that constant curvature solutions of Eq. (5) correspond to the extrema of the potential (13), $V'(\phi)=0$.
The plots of the functions $V(R)$ and $V(\phi)$ (13) are given in Figs. \ref{fig.2} and 3, correspondingly.
\begin{figure}[h]
\includegraphics[height=2.5in,width=2.9in]{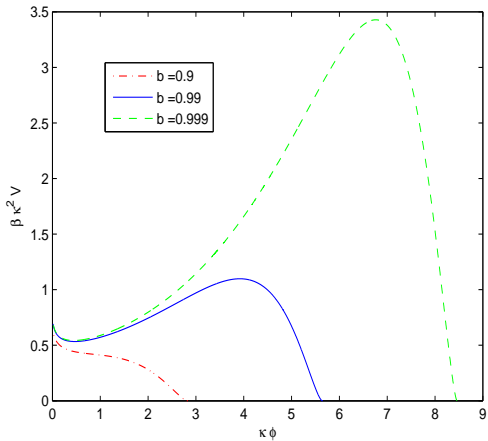}
\caption{\label{fig.3}$\beta \kappa^2 V$ versus $\kappa \phi$.}
\end{figure}
In Fig. 3 the $\kappa \phi$ intercepts correspond to $R=0$. It is clear from Fig. 3 that the potential function $V(\phi)$ has minima at $\kappa \phi_2\approx 0.466$ ($\beta R_2\approx 1.4582$) for $b=0.99$ and $\kappa \phi_2\approx 0.445$ ($\beta R_2\approx 1.5094$) for $b=0.999$, and maximums at $\kappa \phi_1\approx 3.919$ ($\beta R_1\approx 0.1791$) for $b=0.99$ and $\kappa \phi_1\approx 6.760$ ($\beta R_1\approx 0.05495$) for $b=0.999$.
Thus, nontrivial constant curvature solutions of Eq.(5) match the extrema of the potential function (13).
The mass squared of a scalaron is defined by
\begin{widetext}
\begin{equation}
m_\phi^2=\frac{d^2V}{d\phi^2} =\frac{1}{3}\left(\frac{1}{F''(R)}+\frac{R}{F'(R)}-\frac{4F(R)}{F'^2(R)}\right)
=\frac{1+(\beta R)^2}{3\beta}\left[\frac{1+(\beta R)^2}{2b\beta R}+\frac{(1+(\beta R)^2)
\left[4b\arctan(\beta R)-3\beta R\right]-b\beta R}{(1-b+(\beta R)^2)^2}\right].
\label{14}
\end{equation}
\end{widetext}
If the curvature is close to zero the value $m_\phi^2$ is positive and corresponds to a stabile state for $b=0.99$. The de Sitter solutions $\beta R_1\approx 0.1791$ (for $b=0.99$) and $\beta R_1\approx 0.05495$ (for $b=0.999$) give the negative mass squared $m_\phi^2<0$ and corresponding states are unstable, but solutions $\beta R_2\approx 1.4582$ (for $b=0.99$) and $\beta R_2\approx 1.5094$ (for $b=0.999$) give the positive mass squared $m_\phi^2>0$ and corresponding states are stable. These statements are in agreement with previous conclusions made from the analysis of the potential functions. It follows from Eq. (14) that at $\beta R<0.05$ and at $\beta R>0.37$ ($b=0.99$) we have $m_\phi^2>0$; i.e., the space-time is stable. The plots of the function $m^2_\phi$ for different parameters $b$ are given in Fig. \ref{fig.4}.
\begin{figure}[h]
\includegraphics[height=2.5in,width=2.9in]{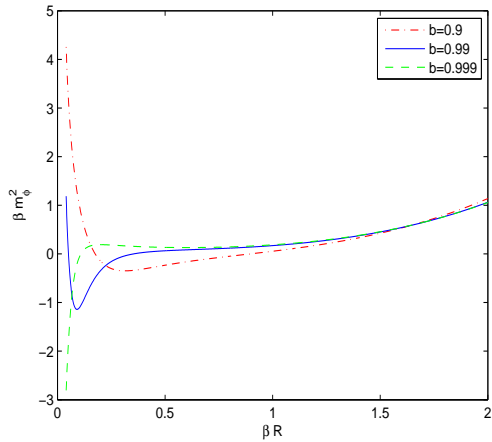}
\caption{\label{fig.4}$\beta m_\phi^2$ versus $\beta R$.}
\end{figure}
The stability of the de Sitter solution in $F(R)$-gravity theory was first discussed in \cite{Schmidt}. As for small values of $\beta R$, $m_\phi^2$ approaches the infinity, and corrections to the Newton law can be ignored. Thus, the Universe in the unstable de Sitter phase inflates and approaches the flat space-time. It is seen from Fig. 2 that
the potential $V(\phi)$ possesses a saddle point at $R=0$; i.e., there is not a local minimum of the function $V(\phi)$ at $R=0$.\\
During the inflation the potential energy of the scalar field is transformed into kinetic energy.
In the models with a minimum of the effective potential, after inflation the Universe was reheated by oscillating the scalar field and elementary particles were created forming the matter of the Universe \cite{Kofman}, \cite{Motohashi}, \cite{Arbuzova}. As a result of damped oscillations, the scalar field transfers its energy to matter. In the Einstein frame the scalar field drives inflation at the early stages of the evolution of the Universe. Thus, reheating takes place due to the particle production, and we come to a state of thermal equilibrium at the reheating temperature. The energy of the classical scalar field is transformed into the thermal energy of elementary particles. Within the inflationary cosmology, after the end of inflation, all particles in the Universe were produced because of quantum effects. Therefore, to describe the process of reheating the Universe precisely one needs to explore the particle theory. But this requires the knowledge of the coupling of the scalar field with other elementary particles. Such complicated studies can be possible by numerical methods. In our model at $R=0$ there is not a minimum of the $V(\phi)$ function, and there are no either free or damped oscillations. Therefore additional investigation is necessary to explain an effective reheating.

\section{Cosmological parameters}

The guarantee that corrections from $F(R)$ gravity compared to GR are small for $R\gg R_0$ ($R_0$ is a curvature at the present time) gives the restrictions \cite{Appleby}
\begin{equation}
\mid F(R)-R\mid \ll R,~\mid F'(R)-1\mid \ll 1,~\mid RF''(R)\mid\ll 1.
\label{15}
\end{equation}
From Eq.(1) we obtain
\[
b\arctan(\beta R) \ll \beta R,~~~~b \ll 1+(\beta R)^2,
\]
\vspace{-7mm}
\begin{equation}
\label{16}
\end{equation}
\vspace{-7mm}
\[
2b(\beta R)^2\ll \left[1+(\beta R)^2\right]^2.
\]
One can verify that at $0<b<1$ all inequalities in Eqs. (16) are satisfied.\\
The slow-roll parameters are given by \cite{Liddle}
\begin{equation}
\epsilon(\phi)=\frac{1}{2}M_{Pl}^2\left(\frac{V'(\phi)}{V(\phi)}\right)^2,~~~~\eta(\phi)=M_{Pl}^2\frac{V''(\phi)}{V(\phi)},
\label{17}
\end{equation}
with the reduced Planck mass $M_{Pl}=\kappa^{-1}$. For the slow-roll approximation we need the conditions $\epsilon(\phi)\ll 1$, $\mid\eta(\phi)\mid\ll 1$. One obtains the slow-roll parameters expressed through the curvature from Eqs. (12)-(14),
\begin{equation}
\epsilon=\frac{1}{3b^2}\left[\frac{\left(1+x^2\right)\left(x-2b\arctan x\right)+bx}{\left(1+x^2\right)\arctan x-x}\right]^2,
\label{18}
\end{equation}
\begin{widetext}
\begin{equation}
\eta=\frac{\left(1+x^2\right)\left[\left(1+x^2\right)^2+8b^2x\arctan x\right]-bx^2(8x^2+10+b)+b(b-2)}{3b^2x\left[\left(1+x^2\right)\arctan x-x\right]},
\label{19}
\end{equation}
\end{widetext}
where $x=\beta R$, and the dependance of curvature $R$ on the inflation field $\phi$ is given by Eq. (12) (see Fig. 1). The plots of the functions $\epsilon$, $\eta$ at $b=0.99$ are given in Figs. 5 and 6, respectively.
\begin{figure}[h]
\includegraphics[height=2.5in,width=2.9in]{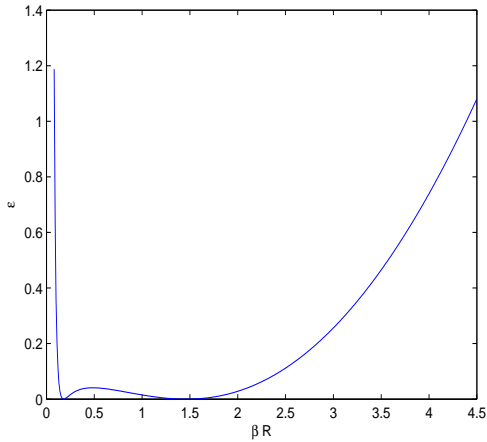}
\caption{\label{fig.5}The function $\epsilon$ versus $\beta R$ ($b=0.99$).}
\end{figure}
\begin{figure}[h]
\includegraphics[height=2.5in,width=2.9in]{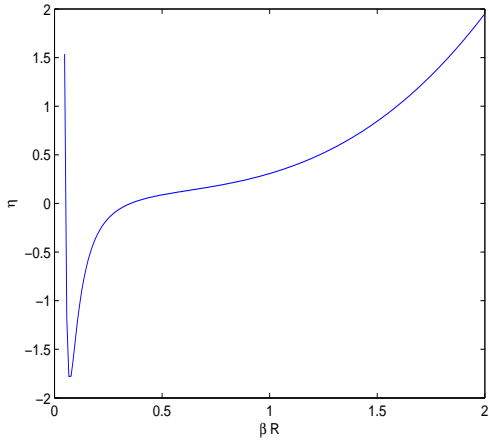}
\caption{\label{fig.6}The function $\eta$ versus $\beta R$ ($b=0.99$).}
\end{figure}
The inequality $\epsilon <1$ holds at $4.39>\beta R>0.083$ and $|\eta | <1$ at $1.59>\beta R>0.12$. As is seen from Figs. 5 and 6 the slow-roll approximation of the model is justified.\\
The age of the inflation may be evaluated by calculating the $e$-fold number \cite{Liddle}
\begin{equation}
N_e\approx \frac{1}{M_{Pl}^2}\int_{\phi_{end}}^{\phi}\frac{V(\phi)}{V'(\phi)}d\phi,
\label{20}
\end{equation}
where $\phi_{\mbox{end}}$ corresponds to the time at the end of inflation. From Eqs. (12), (13), and (20) we obtain the number of $e$-foldings
\begin{widetext}
\begin{equation}
N_e\approx 6b\int_{x_{end}}^x\frac{x-(1+x^2)\arctan x}{x(1+x^2)[x(1+x^2)+bx-2b(1+x^2)\arctan x]}dx.
\label{21}
\end{equation}
\end{widetext}
Here the value $x_{end}=\beta R_{end}$ corresponds to the time of the end of inflation when $\epsilon$ or $|\eta |$ are close to $1$. At $x_{end}=0.18$, $x=1.45$ we get $N_e\approx 63$, which is a reasonable amount of inflation \cite{Liddle}.\\
The index of the scalar spectrum power law due to density perturbations is given by \cite{Liddle}
\begin{equation}
n_s=1-6\epsilon+2\eta.
\label{22}
\end{equation}
With the help of Eqs. (18), (19, and (22), we represent the function of $n_s$ versus $\beta R$ at $b=0.99$ in Fig. 7.
\begin{figure}[h]
\includegraphics[height=2.5in,width=2.9in]{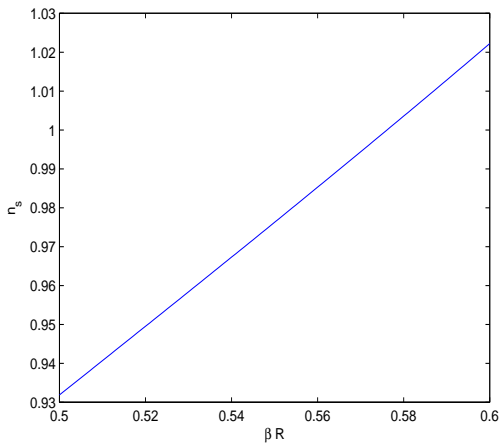}
\caption{\label{fig.7}The function $n_s$ versus $\beta R$ ($b=0.99$).}
\end{figure}
The tensor-to-scalar ratio is defined by \cite{Liddle} $t_s=16\epsilon$. The PLANCK experiment results \cite{Ade} are
\begin{equation}
n_s=0.9603\pm 0.0073,~~~~t_s<0.11.
\label{23}
\end{equation}
From Eq.(22) (see Fig.7) we find that the experimental value of $n_s$ and the inequality $t_s<0.11$ are  satisfied at the curvature $\beta R\approx 0.5321$. Therefore, the model suggested can give a description of inflation.

\section{Critical points and stability}

Following \cite{Amendola} we consider the dimensionless parameters
\begin{equation}
x_1=-\frac{\dot{F}'(R)}{HF'(R)},~x_2=-\frac{F(R)}{6F'(R)H^2},~x_3=\frac{\dot{H}}{H^2}+2,
\label{24}
\end{equation}
\begin{equation}
m=\frac{RF''(R)}{F'(R)},~~~~r=-\frac{RF'(R)}{F(R)}=\frac{x_3}{x_2},
\label{25}
\end{equation}
where the dot over the variable means the derivative on the cosmic time. Then equations of motion can be represented as autonomous equations \cite{Amendola}. To investigate the critical points for the system, one has to study the function $m(r)$ \cite{Amendola} characterizing the deviation from the $\Lambda$CDM model. From Eqs. (1), (3), and (25) we obtain equations as follows
\[
m=\frac{2bx^2}{(1+x^2)(1-b+x^2)},
\]
\vspace{-7mm}
\begin{equation}
\label{26}
\end{equation}
\vspace{-7mm}
\[
r=-\frac{x(1-b+x^2)}{(1+x^2)(x-b\arctan x)},
\]
where $x=\beta R$. The plot of the multivalued function $m(r)$ is given in Fig. 8.
\begin{figure}[h]
\includegraphics[height=2.5in,width=2.9in]{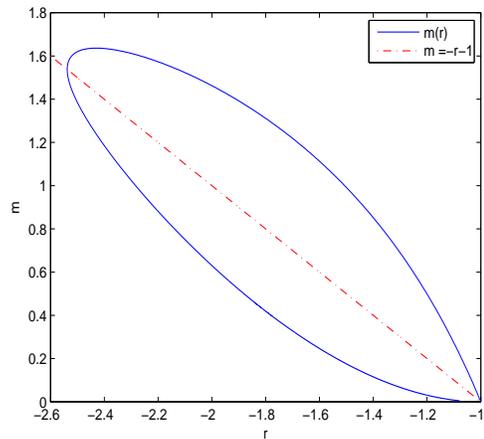}
\caption{\label{fig.8}The function $m$ versus $r$ ($b=0.99$).}
\end{figure}
When the scalar curvature $R$ increases from infinity to zero, the point on the $m(r)$ curve in Fig. 8 moves clockwise in the loop from the point $r=-1$, $m=0$. Let us consider the de Sitter point $P_1$ \cite{Amendola} in the absence of radiation ($x_4 = 0$): $x_1=0$, $x_2=-1$, $x_3=2$. One can verify with the help of Eqs. (1), (3), and (24) that this point corresponds to the constant curvature solutions (6) $x \approx 0.1791$ and $x \approx 1.4582$ ($b=0.99$, $r=-2$) with the matter energy fraction parameter $\Omega_{\mbox{m}}=1-x_1-x_2-x_3=0$ and the effective equation of state (EoS) parameter $w_{\mbox{eff}}=-1-2\dot{H}/(3H^2)=-1$ ($\dot{H}=0$) corresponding to DE. The point $x \approx 0.1791$ ($b=0.99$) belongs to the upper branch of the function $m(r)$, and according to Fig. 8 we have $1< m(r=-2)$. Thus, the condition for the stability of the de Sitter point $0\leq m(r=-2)\leq 1$ \cite{Amendola} is not satisfied. But the point $x \approx 1.4582$ ($b=0.99$) belongs to the lower branch of the function $m(r)$, and one has $0< m(r=-2)<1$ with the stability of this de Sitter point. To have a viable matter dominated epoch prior to late-time acceleration one needs solutions with EoS of a matter era $w_{\mbox{eff}}=0$ ($a=a_0t^{2/3}$ corresponding to $x_3=1/2$) which is realized for the critical point $P_5$ with $m\approx 0$, $r\approx -1$ \cite{Amendola}. The points $P_5$ and $P_6$ belong to the equation $m=-r-1$. The equation $m(r)=-r-1$ with the help of Eqs.(26) becomes
\begin{equation}
\frac{x(3x^2+1-b)}{(1+x^2)^2+b(x^2-1)}=\arctan x.
\label{27}
\end{equation}
Equation (27) possesses two solutions: the trivial solution $x=\beta R=0$ ($m= 0$, $r= -1$) corresponding to the critical point $P_5$ and the nontrivial solution. The nontrivial numerical solution to Eq. (27) for $b=0.99$
gives $x\approx 0.48$, $m\approx 1.54$, and $r\approx -2.54$ (see Fig. (8)). As $m>(\sqrt{3}-1)/2$, this point
belongs to the curvature-dominated point $P_6$ \cite{Amendola} and does not correspond to the standard matter era. The effective EoS becomes $w_{\mbox{eff}}=(2-5m-6m^2)/3m(1+2m)\approx -1.06$, and it satisfies the condition for acceleration. This point is stable and slightly phantom because $w_{\mbox{eff}} <- 1$, $m'(r)>-1$ (see Fig. 8) and corresponds to the region (C) in Fig. 1 of the work \cite{Amendola}. From Eqs.(26) we obtain the derivative $m'(r)=(dm/dx)(dx/dr)$ expressed via $x$,
\begin{widetext}
\begin{equation}
m'(r)=\frac{4x\left(1-b-x^4\right)\left(x-b\arctan x\right)^2}{\left(1-b+x^2\right)^2\left\{x\left(b-1-3x^2\right)+\left[(1+x^2)^2+b(x^2-1)\right]\arctan x\right\}}.
\label{28}
\end{equation}
\end{widetext}
In accordance with Eq. (28) the function $m'(r)$ has singularity at $x=0.1$ and $x\approx 0.48$  ($m\approx 1.54$, $r\approx -2.54$, see also Fig. 8) for $b=0.99$ because the tangent line for these points becomes vertical, $m'=\infty$. When $x$ is the solution of Eq. (27) [$m(r)=-r-1$] the denominator of Eq.(28) is zero and $m'(r)=\infty$.  For the point $x=0$, we have $m=0$, $r=-1$, $m'(r=-1)=0>-1$ and $w_{\mbox{eff}}=0$, $\Omega_m=1$, $a=a_0t^{2/3}$ which are the necessary conditions for the existence of the standard matter era \cite{Amendola}. Thus, the trajectory for which $m$ is close to $0$ and goes near the critical point $P_5$ ($m=0$, $r=-1$) can give acceptable cosmology and be consistent with observations. Then the radiation point $P_8$ exists ($m\approx 0$, $a\propto t^{1/2}$) connecting with the matter point $P_5$ ($m\approx 0$) and the trajectory possibly leads for the late-time acceleration.
There are three general conditions \cite{Amendola} for a successful $F(R)$ model. The first two conditions that are satisfied in our model are the following:
(i) $F(R)$ model has a standard matter dominated epoch only if it satisfies the conditions
$m(r)\approx +0$ and $m'(r)>-1$ at $r\approx -1$.
(ii) The matter epoch is followed by a de Sitter acceleration ($w_{\mbox{eff}}=-1$) only if
$0\leq m(r)\leq 1$ at $r =-2$ (Class II).
The second condition corresponds to the de Sitter state with $x=1.4582$ ($b=0.99$). It should be mentioned that the Class II (as well as the Class IV) model leads to an acceptable cosmology \cite{Amendola}.
For a detailed description of a universe evolution a numerical analysis is necessary to examine the conditions for a late-time accelerator attractor.

\subsection{The effective gravitational constant}

In the modified $F(R)$-gravity theory the effective gravitational constant, related to the evolution of the matter density perturbations, is given by (in our notations) \cite{Tsujikawa}
\begin{equation}
G_{\mbox{eff}}=\frac{G}{F'(R)}\frac{1+4\frac{k^2}{a^2R}m}{1+3\frac{k^2}{a^2R}m},
\label{29}
\end{equation}
where $k$ is a comoving wave number and the parameter $m$ is defined by Eq.(25). On the scales when
large-scale structure is formed the condition $k^2m/(a^2R)\ll 1$ is realized and Eq.(29) becomes \cite{Tsujikawa}
\begin{equation}
G_{\mbox{eff}}=\frac{G}{F'(R)}\left(1+\frac{k^2m}{a^2R}\right).
\label{30}
\end{equation}
In this case the second term in brackets of Eq. (30) is small as compared with the unit. This condition is satisfied for trajectories that are close to the critical point $P_5$ because $m\approx 0$. Then the matter density perturbation
$\delta_m \propto t^{2/3}$, the gravitational potential $\Phi$=const \cite{Tsujikawa}, and we come to the standard result. It should be mentioned that at the present epoch the local gravity constraint requires the equality $k^2m/(a^2R)\ll 1$.\\
The entropy $S$ in $F(R)$-gravity models is given by \cite{Akbar}
$S =F'(R)A/(4G)$, which is the generalization of the Bekenstein-Hawking formula, where $A$ is the area of the horizon. One can obtain the standard formula for the entropy introducing the effective gravitational coupling $G_{\mbox{eff}}=G/F'(R)=G[1+(\beta R)^2]/[1-b+(\beta R)^2]$, which is in accordance with Eq. (30) at $k^2m/(a^2R)\ll 1$.

\section{Conclusion}

The model of $F(R)$ gravity under consideration mimics a cosmological constant and admits two de Sitter solutions ($P_1$ points) for $r=-2$ with $\beta R_1\approx 0.1791$ ($m>1$, $b=0.99$), which is an unstable state and with $\beta R_2\approx 1.4582$ ($0<m<1$, $b=0.99$), which the stable state. Thus, in the model of gravity suggested the cosmic acceleration arises. There are $P_5$, $P_6$ points that are the solutions of the equation $m(r)=-r-1$. The critical point with $r\approx -2.54$, $m\approx 1.54$ ($\beta R\approx 0.48$) does not give the standard matter era. The second critical $P_5$ point corresponds to the standard matter era as $m=0$, $r=-1$, $m'(r)>-1$ and $a= a_0t^{2/3}$. The model suggested belongs to Class II \cite{Amendola}: the $m(r)$ curve connects the vicinity of the point ($r$,$m$)=($-1,0$) to the point $P_1$ ($\beta R_2\approx 1.4582$) located at $r=-2$ ($0<m<1$). Thus, it was shown that critical points $P_1$ and $P_5$, in the classification of autonomous equations \cite{Amendola}, are realized in the model under consideration and the inflation in this version of $F(R)$ gravity is possible. As a result, the model can be observationally acceptable.
GR may be an approximation that describes the Universe at the intermediate cosmic time. If $\beta R\rightarrow \infty$, action (2) approaches the Einstein-Hilbert action with the cosmological constant. It was shown that the slow-roll approximation holds and the cosmological parameters calculated are in agreement with the observed CMB data by the WMAP and PLANCK experiments. To explain all the cosmological periods, from inflation to current accelerated expansion, and to check the viability of the model, one needs to solve and analyze the autonomous equations \cite{Amendola}. We leave such investigation for further study.


\begin{thebibliography}{99}

\bibitem{Guth} A. H. Guth, Phys. Rev. \textbf{D 23}, 347 (1981).

\bibitem{Frieman} J. A. Frieman, M. S. Turner and D. Huterer, Annu. Rev. Astron. Astrophys. \textbf{46}, 385 (2008) (arXiv:0803.0982).

\bibitem{Linde} A. D. Linde, \textit{Particle physics and inflationary cosmology}
  (Harwood Academic Publishers, Chur, Switzerland, 1992).

\bibitem{Caldwell} R. R. Caldwell and M. Kamionkowski, Annu. Rev. Nucl. Part. Sci. \textbf{59}, 397 (2009) (arXiv:0903.0866 [astro-ph.CO]).

\bibitem{Faraoni} T. P. Sotiriou and V. Faraoni, Rev. Mod. Phys. \textbf{82}, 451 (2010) (arXiv:0805.1726).

\bibitem{Capozziello} S. Capozziello and V. Faraoni, \textit{Beyond Einstein Gravity: A Survey of Gravitational
Theories for Cosmology and Astrophysics} (Springer Science+Business Media B.V., New York, 2011).

\bibitem{Odintsov} S. Nojiri and S. D. Odintsov, Phys. Rep. \textbf{505}, 59 (2011)
(arXiv:1011.0544 [gr-qc]).

\bibitem{Starobinsky} A. A. Starobinsky, Phys. Lett. \textbf{B 91}, 99 (1980).

\bibitem{Hu} W. Hu and I. Sawicki, Phys. Rev. \textbf{D 76}, 064004 (2007) (arXiv:0705.1158 [astro-ph]).

\bibitem{Appl} S. A. Appleby and R. A. Battye, Phys. Lett. \textbf{B 654}, 7 (2007) (arXiv:0705.3199).

\bibitem{Star} A. A. Starobinsky, JETP Lett. \textbf{86}, 157 (2007) (arXiv:0706.2041).

\bibitem{Deser} S. Deser and G. W. Gibbons, Classical and Quantum Gravity \textbf{15},
L35 (1998) (arXiv:hep-th/9803049).

\bibitem{Nojiri} S. Nojiri and S. D. Odintsov, Phys. Rev. {\bf D 68}, 123512 (2003)
  (hep-th/0307288).

\bibitem{Carroll} S. M. Carroll, V. Duvvuri, M. Trodden and M. S. Turner, Phys. Rev. \textbf{D 70}, 043528 (2004) (arXiv:astro-ph/0306438).

\bibitem{Amendola} L. Amendola, R. Gannouji, D. Polarski, and S. Tsujikawa, Phys. Rev. \textbf{D 75}, 083504 (2007) (gr-qc/0612180).

\bibitem{Cognola} G. Cognola, E. Elizalde, S. Nojiri, S. D. Odintsov, L. Sebastiani and S. Zerbini, Phys. Rev. \textbf{D 77}, 046009 (2008) (arXiv:0712.4017 [hep-th]).

\bibitem{Linder} E. V. Linder, Phys. Rev. \textbf{D 80}, 123528 (2009) (arXiv:0905.2962 [astro-ph.CO]).

\bibitem{Ferreira} M. Ba\~{n}ados and P. G. Ferreira, Phys. Rev. Lett. \textbf{105}, 011101 (2010) (arXiv:1006.1769 [astro-ph.CO]).

\bibitem{Pani} P. Pani, V. Cardoso, and T. Delsate, Phys. Rev. Lett. \textbf{107}, 031101 (2011) (arXiv:1106.3569 [gr-qc]).

\bibitem{Kruglov} S. I. Kruglov, Int. J. Theor. Phys. \textbf{52}, 2477 (2013) (arXiv:1202.4807 [gr-qc]).

\bibitem{Kruglov1} S. I. Kruglov, Int. J. Mod. Phys. \textbf{A 28}, 1350119 (2013) (arXiv:1204.6709 [gr-qc]).

\bibitem{Barrow} J. D. Barrow and A. C. Ottewill, J. Phys. \textbf{A 16}, 2757 (1983).

\bibitem{Schmidt} V. M\"{u}ller, H.-J. Schmidt, and A. A. Starobinsky, Phys. Lett. \textbf{B 202}, 198 (1988).

\bibitem{Sokolowski} G. Magnano and L. M. Sokolowski, Phys. Rev. \textbf{D 50},
5039 (1994) (arXiv:gr-qc/9312008).

\bibitem{Kofman} L. Kofman, A. D. Linde and A. A. Starobinsky, Phys. Rev. \textbf{D 56}, 3258 (1997)
(arXiv:hep-ph/9704452).

\bibitem{Motohashi} H. Motohashi and A. Nishizawa, Phys. Rev. \textbf{D 86}, 083514 (2012).
(arXiv:1204.1472 [astro-ph.CO]).

\bibitem{Arbuzova} E. V. Arbuzova, A. D. Dolgov, and L. Reverberi, J. Cosmol. Astropart. Phys. \textbf{02}, 049 (2012)
 (arXiv:1112.4995 [gr-qc]).

\bibitem{Appleby} S. A. Appleby, R. A. Battye, and A. A. Starobinsky, J. Cosmol. Astropart. Phys. \textbf{1006}, 005 (2010) (arXiv:0909.1737 [astro-ph.CO]).

\bibitem{Liddle} A. R. Liddle and D. H. Lyth, \textit{Cosmological Inflation and Large-scale Structure},
(Cambridge University Press, Cambridge, UK, 2000).

\bibitem{Ade} P. Ade et al. (Planck Collaboration), arXiv:1303.5083, arXiv:1303.5082, arXiv:1303.5076.

\bibitem{Tsujikawa} S. Tsujikawa, Phys. Rev. \textbf{D 76}, 023514 (2007)
(arXiv:0705.1032).

\bibitem{Akbar} M. Akbar and R.-G. Cai, Phys. Lett. \textbf{B 635}, 7 (2006) (arXiv:hep-th/0602156).

\end{thebibliography}
\end{document}